\documentclass[twocolumn,aps,prl,amssymb,showpacs]{revtex4}
\usepackage[dvips,final]{graphicx}
\begin{document}
\title{On the origin of the electric carrier concentration in graphite} %%
\author{A. Arndt, D. Spoddig}
\author{P. Esquinazi}\email{esquin@physik.uni-leipzig.de}
\author{J. Barzola-Quiquia}
\author{S. Dusari}
\author{T. Butz} \affiliation{Institut f\"ur Experimentelle Physik
II, Universit\"{a}t Leipzig, Linn\'{e}stra{\ss}e 5, D-04103 Leipzig,
Germany}

\begin{abstract}
We investigate the dependence of the electrical resistivity and
magnetoresistance of single crystalline micrometer-sized graphite
samples of a few tens of nanometers thick on the defect concentration
produced by irradiation at low fluences. We show that the carrier
density of graphite $n$ is extremely sensitive to the induced defects
for concentrations as low as $\sim 0.1~$ppm and follows $n \sim
1/R_V^2$ with $R_V$ the distance between defects in the graphene
plane. These and Shubnikov-de Haas oscillations results indicate that
at least a relevant part of the carrier densities measured in
graphite is not intrinsic.
\end{abstract} \pacs{73.90.+f,61.80.-x,81.05.Uw} \maketitle

%\section{Introduction}

The electronic properties of {\em ideal} graphite are actually not
well known simply because {\em defect-free} graphite samples do not
exist. In the last fifty years scientists flooded the literature with
reports on different kinds of electronic measurements on graphite
samples, providing evidence for carrier (electron plus hole)
densities per graphene layer at low temperatures $n_0 \sim 10^{10}
\ldots 10^{12}~$cm$^{-2}$, see e.g. Refs.
\onlinecite{mcc64,kelly,gru08}.   Taking into account that: (1) an
exhaustive experience accumulated in gapless semiconductors - whose
density of states should be similar to its counterpart in a semimetal
- indicates so far that the measured $n_0$ is most probably due to
impurities \cite{tsi97}, and (2) the expected sensitivity of $n_0$ in
graphite on lattice defects and impurities or adatoms
\cite{kelly,sta07}, a fundamental question remains unanswered,
namely, how large is the intrinsic $n_0$ of ideal graphite? Why $n_0$
is so important? Let us recapitulate some fundamental band structure
theoretical results for the graphite structure \cite{kelly}.
Two-dimensional (2D) calculations assuming a coupling $\gamma_0$
between nearest in-plane neighbors C-atoms give a carrier density
(per C-atom)  $n(T) = (0.3 \ldots 0.4) (k_B T/\gamma_0)^2$ ($\gamma_0
\simeq 3~$eV and $T$ is the temperature). Introducing a coupling
($\gamma_1 \sim 0.3~$eV) between C-atoms of type $\alpha$ in adjacent
planes  one obtains $n(T) = a (\gamma_1/\gamma_0^2) T + b
(T/\gamma_0)^2 + c (T^3/\gamma_0^2\gamma_1) + \ldots$ ($a,b,c,\ldots$
are numerical constants). In both cases $n(T \rightarrow 0)
\rightarrow 0$. Neither in single layer graphene nor in  graphite
such $T-$dependences were ever reported \cite{nT}, i.e. a large
density background $n_0$ was always measured and assumed as
``intrinsic" without taking care of any influence from lattice
defects (including edge effects \cite{kob06}) or impurities. To fit
experimental data and obtain a finite Fermi energy $E_F$, up to seven
free parameters were introduced in the past, whereas in the simplest
case $E_F \propto \gamma_2$ \cite{kelly, dil77}.

Clearly, any evidence that speaks against an intrinsic origin of -
even a part of - the measured $n_0$  in graphite samples would cast
doubts on the relevance of related electronic band structure
parameters obtained in the past and will help significantly to
clarify observed transport phenomena. As in the case of gapless
semiconductors \cite{tsi97} this requires a formidable experimental
task. For example, to prove that the measured $n_0 = 2 \times
10^{8}$~cm$^{-2}$ in Ref.~\onlinecite{gar08} is due to
vacancies/interstitials requires a vacancy resolution better than
0.05~ppm. Although nowadays the concentration of impurities in
graphite can be measured with $\sim 0.1~$ppm resolution there is no
experimental method that allows us to determine with such a precision
the number of vacancies or C-interstitials. In spite of that and
because of this situation we would like to start the discussion on
the origin of $n_0$ postulating that at least part of it cannot be
intrinsic. The studies presented here provide answers to: (1) Can a
single vacancy/interstitial provide $\sim$one carrier into the
conduction band even if they are several hundreds of nm apart (ppm
concentration)? This is a relevant issue specially because we expect
that the Fermi wavelength in graphite $\lambda_F \gtrsim 1~\mu$m
\cite{gon07}. (2) Can the resistivity of graphite change with such
small defect concentrations?  (3) How reliable are band structure
parameters of graphite obtained from the field-induced quantum
oscillations in the resistivity (or magnetization)? (4) Why there is
an apparent maximum value for $n_0 \sim 10^{12}~$cm$^{-2}$ in
graphite samples?

In this study we measured the change of the electrical resistance of
thin crystalline graphite samples as a function of defect
concentrations between $\sim 0.1$ to $\sim 10^3$~ppm. To do this we
irradiated three $\sim 60~$nm thick and tens square micrometer
samples under ambient conditions with a focused proton microbeam of
2.25~MeV energy scanned over the samples.  A photo of sample~1 can be
seen in Fig.~1. Particle induced x-ray emission measurements were
done in situ and revealed a total concentration of $\sim 20~\mu$g/g
of non-magnetic impurities except hydrogen of concentration $0.5 \pm
0.3\%$ \cite{hhopg}. The magnetoresistance of a fourth sample of size
$11 \times 2 \times 0.015~\mu$m$^3$ was measured at three different
parts, each of length $\simeq 1.6~\mu$m and irradiated with 30~keV
Ga$^+$ ions. In this sample Shubnikov-de Haas (SdH) oscillations were
measured at 4K before and after irradiation and also for the
corresponding bulk sample. Further details on the electron-beam
spectroscopy and Raman techniques and the exfoliation and ultrasonic
procedures use to characterize and prepare the crystalline thin
graphite flakes from bulk samples will be published elsewhere.

\begin{figure}[]
%\vspace{0.5cm}
\begin{center}
\includegraphics[width=1.0\columnwidth]{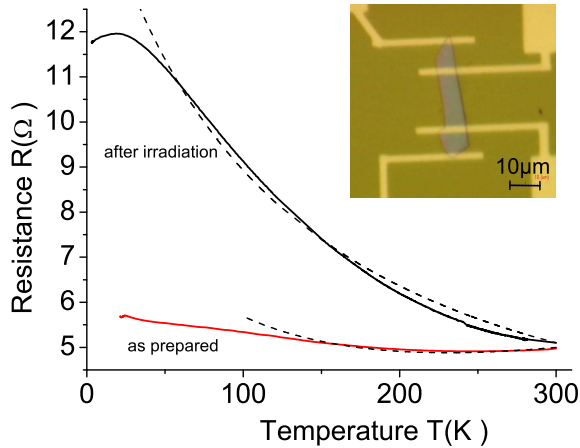}
\caption[]{Temperature dependence of the resistance of sample~3 in
the as-prepared state and after proton irradiation with a fluence of
$9 \times 10^{13}~$cm$^{-2}$ (continuous lines). Note the large
change in $R(T)$ after inducing only $\sim 3~$ppm vacancy density.
The dashed lines are obtained assuming a Fermi energy $E_F = E_0 +
k_B T$ and a $T^{-2}$ dependence for the mean free path, see text for
details. The inset shows an optical microscope picture of sample~1
with gold electrodes on top.} \label{1}
\end{center}
\end{figure}

\begin{figure}[]
%\vspace{0.5cm}
\begin{center}
\includegraphics[width=0.75\columnwidth]{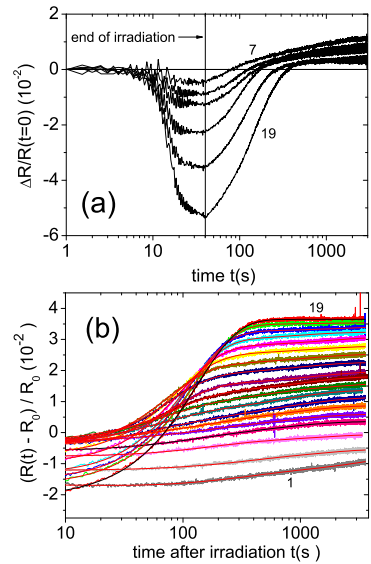}
\caption[]{(a) Relative change of the resistance $R(t) - R(0) / R(0)$
measured vs. time during and after the proton beam hits sample~3.
 The different curves were obtained at different starting
 relaxed conditions after irradiation
of $3.1, 4.0, 4.3, 6.4, 7.2, 9.0 \times 10^{13}$ protons per
cm$^{-2}$, corresponding to the curve numbers 7, 9, 10, 14, 16, 19,
from top to bottom. The small oscillations in the resistance observed
during irradiation are an artifact due to the overlapping of the
proton current and the ac current of the resistance bridge. (b)
Change of resistance relative to its value in the virgin state $R_0$
vs. time. The time scale is taken from the time at which the beam
does not hit the sample anymore, i.e. the minimum in $\Delta R/R(0)$
in (a). Note the decrease of resistance with irradiation for the
first three curves (1-3 from bottom) even in the relaxed states
(after 1 hour). As in (a) the different curves are taken from the
sample at different initial states irradiated with fluences $(1, 2,
..., 19) = (1, 1.5, 2.0, ..., 9) \times 10^{13}$ protons/cm$^{-2}$.}
\label{2}
\end{center}
\end{figure}

Figure~2(a) shows the relative change of the resistance vs. time
during and after irradiation of sample~3 at 297~K. The curves are
obtained at different initial relaxed states after application of a
certain proton fluence. When the beam starts to hit the sample we
observe a clear decrease in the resistance, whose amount depends on
the fluence used and on the initial sample state.  Figure~2(b) shows
the resistance change relative to the sample virgin state. In this
figure $t = 0~$s means the time at which the beam stops irradiating
the sample. Remarkable is that for all samples in the virgin state we
observe a {\em decrease} in the resistance of a few percent for
induced defect density $< 3~$ppm (average defect distance $R_V >
100~$nm!) that remains after several hours after irradiation, i.e. in
the relaxed state $R(t \gtrsim 1$h), see curves (1-3) in Fig.~2(b).
In Fig.~3 we show this relative change for samples~2 and 3. A further
increase of the fluence increases the resistance in the relaxed
state, see Fig.~3. The explanation for this behavior is that defects
increase the carrier density $n$, as theoretically suggested
\cite{sta07}. Because defects also act as scattering centers, both
the carrier mean free path $l(R_V)$ and $n(R_V)$ have to be taken
into account.

We assume graphite as a structure composed of weakly-coupled graphene
sheets \cite{yakadv07}. Within a factor of two the initial value at
297~K before irradiation for the carrier density is $n_i \sim 6
\times 10^{10}~$cm$^{-2}$ and for the mean free path $l_i \sim 50~$nm
\cite{gar08}. The smallness of $l$ at 297~K in comparison to the
sample size allows us to use the Boltzmann-Drude semiclassical
approach. This is important because for our sample sizes and at $T
\lesssim 150~$K there is no straightforward theoretical approach that
includes ballistic and diffusive scattering that allows us to obtain
in a simple way $n(R_V)$ from the resistance. Hall effect
measurements are not necessarily preferred to obtain $n(R_V)$ since:
- the Hall signal depends on at least six unknown parameters ($n, l,
m^\star$ for electrons and holes independently) that change with
defect concentration and $T$; - conventional multiband approaches
appear to be inadequate for graphite \cite{gar08}; - added to these
difficulties, the Hall signal of graphite can be anomalous at $T <
150~$K \cite{kopeah}.

Following SRIM simulations  \cite{ziegler} the produced defect
concentration at a proton fluence of $10^{13}~$cm$^{-2}$ would be
$n_V \sim 10^9~$cm$^{-2}$. Assuming that each defect in the graphene
plane increases by one the carrier number, the increase in carrier
density after irradiating such fluence will be $n_V \sim 10^{-2}
n_{i}$, or in terms of the related wave vector $k_{V} \simeq (\pi
n_V)^{1/2} = 5.6 \times 10^4~$cm$^{-1} \simeq 0.1 k_F$. This simple
estimate reveals that such small defect concentrations are relevant
for the transport.

\begin{figure}[]
%\vspace{0.5cm}
\begin{center}
\includegraphics[width=0.9\columnwidth]{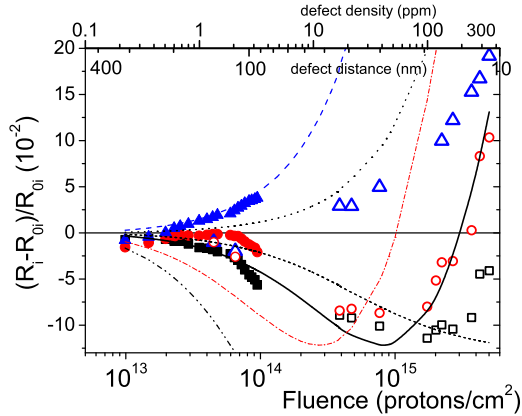}
\caption[]{Relative change of resistance vs. fluence for sample~2
(open symbols) and sample~3 (close symbols). The upper x-axis shows
the corresponding scales as average defect distance within a graphene
plane $R_V$ and the defect density in parts per million (ppm). The
triangles represent the change of the resistance in the relaxed state
after irradiation relative to the virgin state $R_0$, i.e. $R(t
\gtrsim 1$h$)/R_0 -1$. The circles are obtained from $R_{min}/R_0 -
1$ and the squares $R_{\min}/R(t \gtrsim 1$h)$ -1$.
 The curves were obtained from Eq.~(1) with the following
 parameters: $n_V = 0.1/R_V^2, l_i = 50~$nm (long dashed); $0.1/R_v^2, 20~$nm
 (dot);  $1/R_V^2, 50~$nm (continuous); $3/R_V^2, 150~$nm (red dash-dot);
 $3/R_V^2, 50~$nm (dash-double dot). The short-dash curve was obtained
 assuming the usual 3D relation $R \propto 1/ln$ and with $n_V = 0.9/R_V^2, l_i =
 150~$nm. Note that with this 3D
 relationship no minimum in the measured range is obtained within a
 broad variation
 of parameters.} \label{3}
\end{center}
\end{figure}
\begin{figure}[]
%\vspace{0.5cm}
\begin{center}
\includegraphics[width=0.9\columnwidth]{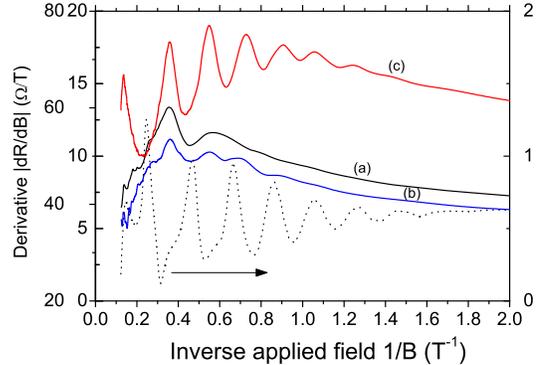}
\caption[]{First derivative of the magnetoresistance measured at 4~K
in two parts (a) and (b) of similar area $2 \times 1.6~\mu$m$^2$
separated by $3~\mu$m in sample~4 ($0-20~\Omega/$T range). (c) Sample
part (a) after irradiation with a fluence of $5 \times
10^{11}~$Ga$^+$-ions per cm$^{2}$ ($20-80~\Omega/$T range). The
dashed curve was obtained for the bulk sample of size $2 \times 1
\times 0.2~$mm$^3$ at 4~K. The magnetic field was applied normal to
the graphene layers.} \label{4}
\end{center}
\end{figure}

At 297~K  the produced defects by irradiation are metastable
\cite{met}, see Fig.~2. Therefore,  we plot in Fig.~3 the relative
change of the resistance just at the end of the irradiation $R_{min}$
with respect to the virgin state
 $R_0$, i.e. $R_{min}/R_0 - 1$ (close and open
circles in Fig.~3). Another possibility is to plot the relative
change with respect to the resistance taken one hour after
irradiation $R_{rel}$, i.e. $R_{min}/R_{rel} - 1$ (close and open
squares in Fig.~3). Both ways  minimize the influence of annealing
effects and provide a similar behavior. These relative changes
indicate that the resistance reaches a minimum  $\sim 90\%$ of its
initial value  at  $R_V = 20 \ldots 30~$nm, see Fig.~3. At higher
fluences the resistance increases because the decrease of $l$ starts
to overwhelm the increase in $n$.

A quantitative description of these data can be done taking into
account the two dimensional resistivity \cite{nom06} $\rho =
2/e^2v_F^2N(E_F)\tau_F$, where $v_F, E_F, \tau_F$ are the Fermi
velocity, the Fermi energy and the scattering relaxation time at
Fermi energy. Using $N(E_F)$ for clean graphene \cite{sta07}, the
expression for $E_F = \hbar v_F k_F$ and $\tau_F = l/v_F$, one
arrives at the simple expression $\rho = (\pi/2)^{1/2} (\hbar/e^2)
l^{-1}n^{-1/2}$. Furthermore, the carrier density increases as $n =
n_i + n_V$, where $n_V =1/R_V^2$ for one carrier per defect.
Following Mathiessen's rule, the mean free path is given by $l^{-1} =
l_i^{-1} + l_V^{-1}$, where $l_i$ is the initial value due to all
scattering centers before irradiation and $l_V$ the mean free path
due to the produced defects. The relative change of resistance can be
written as
\begin{equation}
\frac{R - R_{0}}{R_{0}} = \left (\frac{1}{1 + (n_V/n_i)} \right
)^{1/2} (1 + \frac{l_i}{l_V}) - 1 \,. \label{delta}
\end{equation}
The solid curve shown in Fig.~3 is obtained with $l_V = 1.15 \times
10^6 [$cm$^{-1}]R_V^2[$cm$^{2}]$ for $l_i = 50~$nm and $n_V =
1/R_V^2$. Within logarithmic corrections, the obtained $l_V(R_V)$
function agrees quantitatively with that found in
Ref.~\onlinecite{sta07}. In Fig.~3 we show also other curves obtained
using other values for $l_0$ and pre-factors for $n_V(R_V)$ as well
as assuming the usual 3D relationship $R \propto 1/n$ instead of
$1/n^{1/2}$. The comparison indicates that within a factor of two
$n_V$ is indeed given by $1/R_V^2$ (for $R_V > 10~$nm) and that the
usual 3D relationship for $R$ cannot describe the observed behavior
within a reasonable range of  parameters.

The remarkable increase in $R(T < 300~$K) and the observed change in
the temperature dependence of graphite after inducing only $\sim
3~$ppm defect density ($R_V \sim 100~$nm) is mainly given by the
decrease of $E_F \simeq E_F(0) + k_B T \propto \sqrt{n(T)}$ with
temperature. As shown in Ref.~\onlinecite{gar08} $E_F$ is basically
determined by thermal electrons (note that $E_F \sim 330~$K for $n =
6 \times 10^{10}~$cm$^{-2}$) and its $T-$dependence overwhelms that
of $l(T)$. With $E_F(T)$, $l_i(T) \propto T^{-2}$~\cite{gar08} and
the parameters obtained from Fig.~3 in Eq.~(1) one can understand the
observed temperature dependence of the resistance above $\sim 100~$K,
see Fig.~1; at lower temperature the Boltzmann-Drude approach looses
its validity.

The band parameters of graphite were obtained {\em mostly} on
macroscopic samples and usually from magneto-optical studies, SdH and
de Haas-van Alphen oscillations, cyclotron resonance, etc.  We doubt
that in graphite samples the defect density is negligible and
therefore we expect that the carrier density is neither small nor
homogeneously distributed. Within the $11~\mu$m length of sample~4 we
measured the magnetoresistance at 4~K and calculate its first
derivative in different parts of similar area. The SdH oscillations
depend on the sample position, see Fig.~4, indicating clearly
inhomogeneities in the carrier concentration within micrometers in
agreement with EFM results that revealed sub-micrometer domainlike
carrier density distributions in graphite surfaces \cite{lu06}.

For the measured sample area that gives curve (a) in Fig.~4 and
within experimental resolution there are no SdH oscillations up to a
field $B \simeq 1.8$~T in clear contrast to the bulk sample, see
Fig.~4. This fact can be understood assuming that in most of this
sample part $n_0 \lesssim 10^{9}$cm$^{2}$. Then, the corresponding
Fermi wavelength $\lambda_F \gtrsim 0.8~\mu$m is of the order of the
sample size and larger than the cyclotron radius $r_c = m^\star v_F
/eB$ for $B > 0.07~$T assuming $m^\star = 0.01 m$ ($m$ is the free
electron mass). In this case we do not expect to observe any SdH
oscillations. However, for $B \simeq 1.8~T$ and 2.8~T two maxima are
observed. From the measured ``period" $P$ in $1/B$ as well as from
the first field at which the first maximum appears we estimate the
existence of domains of size $ < 2 r_c \lesssim 100~$nm~ in which
$\lambda_F \lesssim 50~$nm, i.e. domains with $n_0 \gtrsim
10^{11}~$cm$^{-2}$ within a matrix of much lower carrier
concentration. This indicates that the description of the SdH
oscillations in real graphite samples can be achieved only within the
framework of inhomogeneous 2D systems \cite{dol96,har01},  an issue
rarely studied in the past.

The selected Ga$^+$ irradiation produced an average defect
concentration of $\sim 10^{12}~$cm$^{-2}$, i.e. $\sim 10^3$~ppm (Ga
implantation $\lesssim 1$~ppm) in the thin graphite sample,
``homogenizing" its carrier density distribution. After irradiation
the SdH oscillations are clearly observed for $B \gtrsim 0.7~$T, see
Fig.~4. Their period
 $0.16~$T$^{-1} \lesssim P \lesssim 0.23~$T$^{-1}$ is
 within the range found in literature \cite{kelly,luky04} and
 indicates $n_0 \sim 3 \times 10^{11}~$cm$^{-2}$.

 Finally, if a relevant part of the
reported carrier concentration in graphite is due to defects, why
does it appear to saturate at $n_0 \sim 10^{12}~$cm$^{-2}$ ($\sim 3
\times 10^{-4}$ carrier per C-atom)~? We note that such saturation is
also observed in gapless semiconductors with increasing donor
concentration \cite{tsi97}. We might therefore expect it when the
average  distance between defects is of the order of the range of
modification of the electronic structure produced by, e.g. a single
vacancy, found experimentally to be $\sim 3~$nm \cite{rufi00}
implying $n_0 \lesssim 10^{13}~$cm$^{-2}$.

Concluding, the obtained results  indicate that a concentration of
defects (or impurities) of $\sim 0.2$~ppm can generate a carrier
density $\sim 10^9~$cm$^{-2}$ affecting the transport properties.
This is an extraordinary sensitivity taking into account the large
distances between defects these concentrations imply. Taking into
account that, in best case, we have an impurity concentration
$\lesssim 20$~ppm, except for hydrogen $\lesssim 1\%$, plus an
unknown concentration of vacancies and interstitials, we should doubt
about the assumed ``intrinsic" origin of the measured carrier
concentrations in graphite. The behavior of SdH oscillations in
micrometer-sized graphite regions and their changes after introducing
defects support the above statement and indicate that real graphite
is composed by an inhomogeneous distribution of carrier density.

\acknowledgements This work has been possible with the support of the
DFG under DFG ES 86/16-1. S.D. is supported by BuildMona. One of the
authors (P.E.) gratefully acknowledge  correspondence with A. V.
Krasheninnikov and F. Guinea.

%\bibliographystyle{apsrev}
%\bibliography{D:/DATA/hopg/magnetic_carbon}
%\bibliography{D:/data/HOPG/magnetic_carbon}

\begin{thebibliography}{23}
\bibitem[{nT()}]{nT}
\bibinfo{note}{Note, however, that the data obtained in
  Ref.~\protect\onlinecite{gar08} indicate $n[$cm$^{-2}] \sim n_0 + 7 \times
  10^5 T^2$ (within factor of two), or $n = n_0 + 10^5 T^2 + 7.5 \times 10^3
  T^3$ with $T$ in [K].}
\bibitem[{met()}]{met}
\bibinfo{note}{A quantitative description of the observed time dependence
  (Fig.~2) shows that this is governed by interstitial migration processes
  \protect\cite{niw95} and single vacancy diffusion \protect\cite{lee05_vac}
  with activation energies $\sim 0.87~$eV and $\sim 0.93~$eV, respectively. The
  migration processes annihilate partially the produced defects and at fluences
  (for sample~3) $\gtrsim 2 \times 10^{13}~$cm$^{-2}$ the relaxed resistance
  increases indicating that the graphite structure remains with a certain
  amount of defects decreasing the overall carrier mobility.}
\end{thebibliography}

\end{document}